\documentclass[12pt]{article}
\begin{document}
\title{{\bf SUBSYSTEM ENTROPY
\\ EXCEEDING BEKENSTEIN'S BOUND}
\thanks{Alberta-Thy-08-00, hep-th/0007237}}
\author{
Don N. Page
\thanks{Internet address:
don@phys.ualberta.ca}
\\
CIAR Cosmology Program, Institute for Theoretical Physics\\
Department of Physics, University of Alberta\\
Edmonton, Alberta, Canada T6G 2J1
}
\date{(2000 July 28)}
\maketitle
\large

\begin{abstract}
\baselineskip 16 pt

	If Bekenstein's conjectured bound
on the microcanonical entropy,
$S \leq 2 \pi E R$,
is applied to a closed subsystem
of maximal linear size $R$
and excitation energy up through $E$,
it can be violated by an arbitrarily large factor
by a scalar field having a symmetric potential
allowing domain walls, and
by the electromagnetic field modes
in a coaxial cable.

\end{abstract}
\normalsize
\baselineskip 16 pt
\newpage

	Motivated by considerations
of the generalized second law of thermodynamics
for processes in which objects are lowered into black holes,
Bekenstein
\cite{Bek1}
conjectured that the entropy $S$
of a system confined to radius $R$
or less and energy $E$ or less
would obey the inequality
 \begin{equation}
 S \leq 2 \pi E R.
 \label{eq:1}
 \end{equation}
He and colleagues found many arguments and examples
supporting this inequality
[1-14],
though many counterarguments and counterexamples
have also been noted
[15-28].

	Part of the difficulty of determining
whether the bound (\ref{eq:1}) is correct or not
is the ambiguity of what systems are to be considered,
and what definitions are to be assigned to $S$, $R$, and $E$.
To give a concrete resolution of the first ambiguity,
here I shall consider only nongravitational quantum field theories
of a single quantum field in Minkowski spacetime,
avoiding counterexamples with large numbers of fields
\cite{Page1,UW1,FMW}
(unless it is assumed that the fields have
enough positive vacuum energy included in $E$
to save the bound
\cite{Bek3,Bek6,Bek7}).

	For the definitions of $S$, $R$, and $E$,
Bekenstein initially
\cite{Bek1,Bek2}
took $S$ to be the von Neumann
entropy $-Tr(\rho\ln\rho)$ of the system,
$R=(A/4\pi)^{1/2}$ with
``area $A$ of that (quasi) spherical surface
which circumscribes the system,''
and $E$ to be the mean regularized energy, $Tr(H\rho)$,
of the ``complete system,'' which was meant to include
any walls needed to keep the system within the
(quasi) spherical surface of area $A=4\pi R^2$.
With these definitions, the truth of the conjectured
bound (\ref{eq:1}) for nongravitational systems
of a suitably small number of quantum fields
in Minkowski spacetime seems to be an open question,
since it is difficult to give a detailed description
of walls that may be needed to confine the system.
If the walls are themselves made up of quantum fields
that are not held in place by yet other walls,
I would conjecture that no complete stationary system can
be totally confined to be within any finite radius $R$
of Minkowski spacetime,
in which case the right hand side of (\ref{eq:1})
is infinite for any system with positive energy,
making the bound trivially true for complete systems
with a nondegenerate vacuum that has nonnegative energy.

	Therefore, to give (\ref{eq:1}) a nontrivial
content, one might prefer to apply it to a subsystem
of the universe, say a quantum field inside some
boundary circumscribed by a sphere of radius $R$.
Indeed, this is an approach that Schiffer and Bekenstein
have advocated as follows
\cite{SB1}:

	``But there is no gainsaying the conceptual clarity
gained when the bound (1.1) [(\ref{eq:1}) above]
is regarded as applying to the field in the cavity
and only the field.  This motivates an alternative
approach to bound (1.1) which abandons the canonical
method in favor of the microcanonical one
(entropy is the logarithm of the number of available
microstates), interprets $E$ as the available energy
above the vacuum state, and ignores the walls of the cavity. \ldots

	``If $\Omega(E)$ denotes the number of quantum states
accessible to the field system with energy up to and including
$E$, then the microcanonical definition of entropy is
$S(E) = \ln{\Omega(E)}$.''

	This approach avoids counterexamples to (\ref{eq:1})
\cite{Unw,Page1,AW}
from negative Casimir energies by simply redefining
the ground state energy of the field in the cavity
to be zero.  By using the microcanonical entropy
$\ln{\Omega(E)}$ instead of the canonical or von Neumann
entropy $-Tr(\rho\ln\rho)$, it also avoids
counterexamples to (\ref{eq:1})
\cite{Deu,FMW,Page2}
from very low temperatures when $E=0$ for the ground state,
as it is defined to be here.

	Using these definitions of $S$, $R$, and $E$,
I now wish to show two types of counterexamples
for the conjectured entropy bound (\ref{eq:1})
for a single quantum field within a cavity.
For each the ratio of the left hand side
to the right hand side,
 \begin{equation}
 B \equiv {S \over 2 \pi E R},
 \label{eq:2}
 \end{equation}
can be made arbitrarily large,
thereby violating the conjectured entropy bound (\ref{eq:1})
by an arbitrarily large factor.

	The first counterexample uses a field with nontrivial
self-interaction
\cite{Page2},
which old arguments
\cite{Page1}
had suggested could be used to violate (\ref{eq:1}).

	In particular, consider a scalar field
$\phi$ whose potential energy density $V(\phi)$
is symmetric in $\phi$ and has its global
minima at $\phi = \pm \phi_m \neq 0$,
and impose the condition $\phi = 0$ at the boundary
of the region under consideration,
say a cavity of radius $R$.
If the region is large enough,
the energy of a classical configuration
with $\phi = 0$ at the boundary will have
a global minimum for a configuration
in which $\phi$ is near $\phi_m$ over most
of the region and then drops smoothly to 0
at the boundary.
(The region must be large enough that
the reduction in the potential energy density
from $V(0)$ to $V(\phi_m)$ integrates to more
than the increase in the ``kinetic''
or gradient energy density from the spatial
gradients of $\phi$ near the boundary.
For a spherical region, the potential energy
reduction is of the order of $[V(0)-V(\phi_m)]R^3$,
whereas the gradient energy increase is of the order
of at least $(\phi_m/R)^2 R^3$,
so the former definitely dominates if
 \begin{equation}
 R \gg \phi_m/\sqrt{V(0)-V(\phi_m)},
 \label{eq:4}
 \end{equation}
allowing a nonuniform $\phi(x^i)$ to minimize the energy.)

	Classically, there is another global
energy minimum of exactly the same energy
with $\phi(x^i)$ replaced by $-\phi(x^i)$.
But quantum mechanically, there will be some
tiny tunneling rate between these two classical
configurations, so the quantum ground state,
of energy $E_0$,
will be a symmetric superposition of the two
classical energy minima (plus quantum fluctuations
of all the other modes).
However, there will also be an excited state
of energy $E_1 > E_0$ which is the antisymmetric
superposition of the two classical energy minima
(plus other fluctuations).
For a large region, the energy excess $E_1-E_0$
of this excited state will be exponentially tiny.
The number of orthogonal quantum states
with energy up to and including $E=E_1$ will be
$\Omega(E_1)=2$, so the microcanonical entropy
of that energy is $S(E_1)=\ln\Omega(E_1)=\ln 2$.

	When the two classical extrema configurations
$\phi(x^i)$ and $-\phi(x^i)$ are well separated,
we can estimate that the excitation energy $E_1 - E_0$
is given by some energy scale multiplied by $e^{-I}$,
where $I$ is the Euclidean action of an instanton
that tunnels between the two classical extrema
configurations $-\phi(x^i)$ and $+\phi(x^i)$.
This instanton will be a solution of the Euclidean
equations of motion of the field $\phi$ that
obeys the boundary condition $\phi = 0$ at
spatial radius $r = R$ for all Euclidean times $\tau$,
but which for $r < R$ interpolates between
$-\phi(x^i)$ and $+\phi(x^i)$ as $\tau$
goes from $-\infty$ to $+\infty$.
When the strong inequality (\ref{eq:4}) applies,
the static configuration $+\phi(x^i)$
that applies asymptotically for large positive $\tau$
is very near $\phi_m$ over almost all of the
spatial volume (except very near $r = R$), and
the Euclidean instanton is essentially a domain
wall concentrated at some Euclidean time that can
be chosen to be $\tau = 0$.

	The energy-per-area or action-per-three-volume
of the domain wall is
 \begin{equation}
 \varepsilon=\int_{-\phi_m}^{+\phi_m}\sqrt{2V(\phi)}\,d\phi,
 \label{eq:5}
 \end{equation}
and the three-volume of the Euclidean section at
$\tau = 0$ across the ball $r \leq R$ is $4\pi R^3/3$,
so the Euclidean action of the instanton is
 \begin{equation}
 I \approx {4\pi\over 3} R^3 \varepsilon.
 \label{eq:6}
 \end{equation}

	A suitable energy scale to multiply $e^{-I}$
is $R^2 \varepsilon$.  At our level of approximation,
it does not help to try to get the numerical
coefficient of the energy scale correct,
since our estimate (\ref{eq:6}) of the Euclidean action,
though being the dominant piece when it is large
in comparison with unity, has smaller corrections
that are also large in comparison with unity.
Therefore, using $\sim$ to mean that the logarithms
of the two sides are approximately equal
(up to differences that are small in comparison
with the logarithms themselves but which may actually
be large in comparison with unity,
so that the ratio of the two sides themselves
may be much different from unity), we get 
 \begin{equation}
 E_1-E_0 \sim R^2 \varepsilon e^{-I}
     \sim R^2 \varepsilon
        \exp{\left(-{4\pi\over 3}R^3\varepsilon\right)}.
 \label{eq:7}
 \end{equation}

	When the strong inequality (\ref{eq:4}) holds,
classically the energy $E_0$ is essentially the energy
of a half-thickness of domain wall at the radius $R$,
approximately $2\pi R^2 \varepsilon$.
As Bekenstein has noted
\cite{Bek9},
for large $I$ this classical energy
(and also the energy of any wall outside $R$
needed to keep $\phi = 0$ at $R$)
dominates over $E_1-E_0$
(though, incidentally, it does not go
to zero as $R \rightarrow \infty$,
as he also claimed).
However, here I am only considering
the field subsystem inside $R$
(and so am not counting the walls),
and I am also following the approach
of Schiffer and Bekenstein
\cite{SB1}
in taking ``$E$ as the available energy
above the vacuum state''
(which avoids counterexamples to
the conjectured bound for other closed
subsystems with negative Casimir energy).

	Therefore, taking $E = E_1 - E_0$ gives
 \begin{equation}
 B \equiv {S \over 2 \pi E R}
   = {\ln{2}\over 2\pi (E_1-E_0) R}
   \sim {e^I \over I}
   \sim {\exp{\left({4\pi\over 3}R^3\varepsilon\right)}
        \over R^3 \varepsilon}
   \sim \exp{\left({4\pi\over 3}R^3\varepsilon\right)},
 \label{eq:8}
 \end{equation}
which can be made arbitrarily large by making $R$
arbitrarily large.  Indeed, $B-1$, the violation
of Bekenstein's conjectured bound (\ref{eq:1}) if
it is positive, grows large very rapidly with $R$
large enough to obey the inequality (\ref{eq:4}).

	For instance, take a toy model in which
 \begin{equation}
 V(\phi) = {\lambda\over 4}(\phi^2-\phi_m^2)^2
         = {\lambda\over 4}\phi^4 - {1\over 4}m^2\phi^2
	     + {m^4\over 16\lambda}
 \label{eq:9}
 \end{equation}
where $\lambda \ll 1$ and $m$
is the mass of small field oscillations
around the potential minima at
$\phi = \pm\phi_m = \pm m/\sqrt{2\lambda}$.
(If it is objected that this model
is actually a trivial quantum field theory
\cite{Cal,Bek9},
then use instead a well-defined supersymmetric model in which
the bosonic sector has a scalar field with this potential.)
Then
 \begin{equation}
 \varepsilon = {2\over 3}\sqrt{2\lambda}\,\phi_m^3
 = {m^3 \over 3\lambda},
 \label{eq:10}
 \end{equation}
so for $R \gg \phi_m/\sqrt{V(0)-V(\phi_m)} =
2\sqrt{2}/m$,
one gets
 \begin{equation}
 I \approx {8\sqrt{2}\,\pi\over 9}\lambda^{1/2}\phi_m^3 R^3
    = {4\pi\over 9\lambda}m^3 R^3 \gg 1.
 \label{eq:11}
 \end{equation}
When this Euclidean tunneling action is inserted
into Eq. (\ref{eq:8}), one gets that
the quantity $B$, which Bekenstein has conjectured
to be bounded above by unity, is
 \begin{eqnarray}
 B & \sim & {e^I \over I} \sim e^I
   \sim \exp{\left({8\sqrt{2}\,\pi\over 9}
     \lambda^{1/2}\phi_m^3 R^3 \right)}
   = \exp{\left({4\pi\over 9\lambda}m^3 R^3 \right)} \nonumber \\
   & \sim \!\!\! & \!\! 10^{10^{100}} \! \exp{\!\left\{ \left( 10^{100}\ln{10} \right)
   \left[ \left( {10^{-12}\over\lambda} \right)
   \left( {m\over 10^{16}\: {\rm GeV}} \right)^3
   \left( {R\over 0.502269{\rm cm}} \right)^3 \! - \! 1 \right] \right\} }.
 \label{eq:12}
 \end{eqnarray}
This is thus larger than a googolplex for
 \begin{equation}
 R > 0.50227 \; {\rm cm}\left({10^{16}{\rm GeV}\over m}\right)
     \left({\lambda\over 10^{-12}}\right)^{1/3}.
 \label{eq:13}
 \end{equation}

	The second counterexample to Bekenstein's
conjectured entropy bound (\ref{eq:1}) in the microcanonical
approach uses simply the free electromagnetic field
inside the annular region of a long coaxial cable
that forms a closed loop of length $L \gg R$
coiled up inside the sphere of radius $R$.

	For simplicity, take the inner and outer coaxial
cable cylinders to provide perfectly conducting
boundary conditions for the electromagnetic field
between them, and let this region between
the perfectly conducting cylinders be vacuum
(except for the electromagnetic field itself),
rather than the dielectric used
to keep the inner and outer conducting cylinders apart
in realistic coaxial cables.
As Schiffer and Bekenstein advocated
\cite{SB1},
we shall again take the energy $E$ to be the available
electromagnetic field energy over that
of the electromagnetic vacuum within the long annular cavity,
ignoring the energy of the walls of the cavity itself
(the hypothetical perfectly conducting cylinders).

	When the coaxial cable lies along the $z$-axis,
and $\rho = \sqrt{x^2 + y^2}$
is the cylindrical radial coordinate,
the transverse electromagnetic (TEM) wave mode
\cite{Jackson75},
in the vacuum annular region $\rho_1 < \rho < \rho_2$
between the inner perfectly conducting cylinder
at radius $\rho_1$
and the outer perfectly conducting cylinder
at radius $\rho_2$,
has an electromagnetic field two-form
 \begin{equation}
 \mathbf{F} = \mathbf{d}[f(t-z)+g(t+z)]
 \wedge \mathbf{d}\ln{\rho},
 \label{eq:14}
 \end{equation}
a radial electric field and azimuthal magnetic field
whose strengths vary as $1/\rho$ in the radial direction
and have a wave behavior in the time and $z$-direction.
Here $f(t-z)$ gives a TEM wave moving at the speed of light
in the positive $z$-direction, and $g(t+z)$ gives
a TEM wave moving at the speed of light in the
negative $z$-direction.

	When the coaxial cable is coiled up
so that it lies entirely within the sphere of radius $R$,
the TEM wave forms will be altered
by a fractional amount that is small
when the radius of bending of the cable is much greater
than its cylindrical radius $\rho_2$, which will be
assumed to be the case here.  Then TEM waves will
still move along the cable with very nearly the
speed of light.  If the two ends of the cable are connected
together to form a closed loop of length $L$,
the angular frequency of TEM modes in the loop
will be very nearly
 \begin{equation}
 \omega_n = 2\pi |n|/L
 \label{eq:15}
 \end{equation}
for each nonzero integer $n$.
(Each eigenfrequency has two modes, $n = \pm |n|$, one for each
direction the wave can go around the loop.)

	Now if we take the energy
$E = \omega_1 = 2\pi/L$,
then $\Omega(E)$, the number of quantum states
accessible to this electromagnetic field system
with energy up to and including $E$,
is 3, since there is the electromagnetic vacuum
state with energy defined to be zero,
the state with one photon of energy $E$
in the mode $n=1$,
and the state with one photon of energy $E$
in the mode $n=-1$
(going in the opposite direction around the loop).
Therefore, the microcanonical definition
of the entropy for this energy is
$S(E) = \ln{\Omega(E)} = \ln{3}$, so
 \begin{equation}
 B \equiv {S \over 2 \pi E R} = {(\ln{3}) L \over 4 \pi^2 R}.
 \label{eq:16}
 \end{equation}
This exceeds unity if 
 \begin{equation}
 L > {4 \pi^2 \over \ln{3}} \, R \approx 35.9348 \, R.
 \label{eq:17}
 \end{equation}

	For example, suppose that we coil up
the coaxial cable of cylindrical radius $r=\rho_2$
into a close-packed roll, so that in a small cross-section
locally perpendicular to the coiled cable, each piece of the
cable intersecting this cross-section takes up
an hexagonal area $A = \sqrt{12} r^2$.
If we fill the sphere of radius $R$
(and hence volume $V = 4\pi R^3/3$)
with these coiled up hexagonal cylinders
that circumscribe the circular cylinders
of the cable itself, so $L A = V$,
then $L = 2\pi R^3/(3\sqrt{3}r^2)$, giving
 \begin{equation}
 B = {\ln{3}\over 6\sqrt{3} \pi} \left({R\over r}\right)^2.
 \label{eq:18}
 \end{equation}
This obviously can be made arbitrarily large
by making the radius $R$ of the sphere
enclosing the electromagnetic field system
arbitrarily larger than the cylindrical
radius $r$ of the outer perfectly
conducting cylindrical boundary
of the coaxial cable loop containing the
electromagnetic field system.

	This example appears to be not only
a counterexample to
Bekenstein's conjectured entropy bound
$B \leq 1$
in the microcanonical form 
for a closed subsystem of maximal linear size $R$,
but also a counterexample to what is claimed
to be proved in
\cite{SB1},
that the microcanonical bound holds
``for a generic system consisting of
a noninteracting quantum field
in three space dimensions confined to a cavity
of arbitrary shape and topology,''
``any free quantum field
which is described by a Hermitian,
positive-definite Hamiltonian $H$.
Simple examples are the scalar,
electromagnetic, and Dirac fields.''

	The question now arises as to
whether Bekenstein's conjectured entropy
bound can be differently interpreted
so that it is still viable.
As Bekenstein has emphasized
\cite{Bek9}
in his rebuttal to an earlier paper
of mine
\cite{Page2}
that gave similar counterexamples
to various forms of his conjecture,
it may be best to restrict the conjecture
to complete systems
\cite{Bek3,Bek6}.
If the conjecture is then to be nontrivial,
the challenge would be to find
a definition of what it means for
a complete system to be circumscribed
by finite radius $R$.  Work on this is
reported elsewhere
\cite{Page3}.

	Some of this work was done in Haiti
while awaiting the adoption papers
for our now 979-day-old daughter
Ziliana Zena Elizabeth.
This research was supported in part by
the Natural Sciences and Engineering Research
Council of Canada.


\end{document}